\documentclass{PoS}

\newcommand{\be}{\begin{equation}}
\newcommand{\ee}{\end{equation}}
\newcommand{\bea}{\begin{eqnarray}}
\newcommand{\eea}{\end{eqnarray}}

\newcommand{\cffour}{CF$_4$}

\newcommand{\simgt}{\lower.5ex\hbox{$\; \buildrel > \over \sim \;$}}
\newcommand{\simlt}{\lower.5ex\hbox{$\; \buildrel < \over \sim \;$}}

\title{DMTPC:  Dark matter detection with directional sensitivity}

\ShortTitle{The DMTPC directional dark matter detector}

\author{\speaker{J.B.R.~Battat},$^a$ 
S. Ahlen,$^b$
T.~Caldwell,$^a$
C.~Deaconu,$^a$ 
D.~Dujmic,$^{a,c}$ 
W.~Fedus,$^a$ 
P.~Fisher,$^{a,c,d}$ 
F.~Golub,$^e$
S.~Henderson,$^a$ 
A.~Inglis,$^b$
A.~Kaboth,$^a$ 
G.~Kohse,$^f$
R.~Lanza,$^f$
A.~Lee,$^a$ 
J.~Lopez,$^a$ 
J.~Monroe,$^a$ 
T.~Sahin,$^a$ 
G.~Sciolla,$^a$ 
N.~Skvorodnev,$^e$
H.~Tomita,$^b$
H.~Wellenstein,$^e$
I.~Wolfe,$^a$ 
R.~Yamamoto,$^a$ 
H.~Yegoryan$^a$\\
        \llap{$^a$}Physics Department, Massachusetts Institute of Technology, Cambridge, MA 02139, USA\\
        \llap{$^b$}Physics Department, Boston University, Boston, MA 02215, USA\\
        \llap{$^c$}Laboratory for Nuclear Science, Massachusetts Institute of Technology, Cambridge, MA 02139, USA\\
        \llap{$^d$}MIT Kavli Institute for Astrophysics and Space Research, Massachusetts Institute of Technology, Cambridge, MA 02139, USA\\
        \llap{$^e$}Physics Department, Brandeis University, Waltham, MA 02453, USA\\
        \llap{$^f$}Nuclear Science and Engineering Department, Massachusetts Institute of Technology, Cambridge, MA 02139, USA\\
        E-mail: \email{jbattat@mit.edu}}

\abstract{The Dark Matter Time Projection Chamber (DMTPC) experiment uses CF$_4$ gas at low pressure (0.1 atm) to search for the directional signature of Galactic WIMP dark matter.  We describe the DMTPC apparatus and summarize recent results from a 35.7 g-day exposure surface run at MIT.  After nuclear recoil cuts are applied to the data, we find 105 candidate events in the energy range $80-200$~keV, which is consistent with the expected cosmogenic neutron background.  Using this data, we obtain a limit on the spin-dependent WIMP-proton cross-section of $2.0\times 10^{-33}$~cm$^2$ at a WIMP mass of 115~GeV/c$^2$.  This detector is currently deployed underground at the Waste Isolation Pilot Plant in New Mexico.}

\FullConference{Identification of Dark Matter 2010-IDM2010\\
		July 26-30, 2010\\
		Montpellier France}

\begin{document}

\section{The Dark Matter Time Projection Chamber (DMTPC) Detector}
The Dark Matter Time Projection Chamber (DMTPC) collaboration has
developed and operated a 10-liter gas-based directional dark matter
detector.  The current instrument consists of a dual TPC, filled with
\cffour~gas at 75~Torr (3.3~g target mass).  Proportional scintillation from the
avalanches is read out with two CCD cameras, allowing reconstruction of the topology of nuclear recoil tracks.  The induced charge on the TPC
anode is also measured.  DMTPC has demonstrated
head-tail sensitivity for neutron-induced recoils above 100~keV, and
an angular resolution for track reconstruction of 15$^\circ$ at
100~keV.  A 20-liter detector has been built and is currently under commissioning.  The 10-liter detector has completed a surface run to study backgrounds and is currently deployed underground at the Waste Isolation Pilot Plant (WIPP -- near Carlsbad, NM) at a depth of 1600 meters water equivalent.

\subsection{Detector description and performance}
\begin{figure}
\centering 
\includegraphics[width=0.6\textwidth]{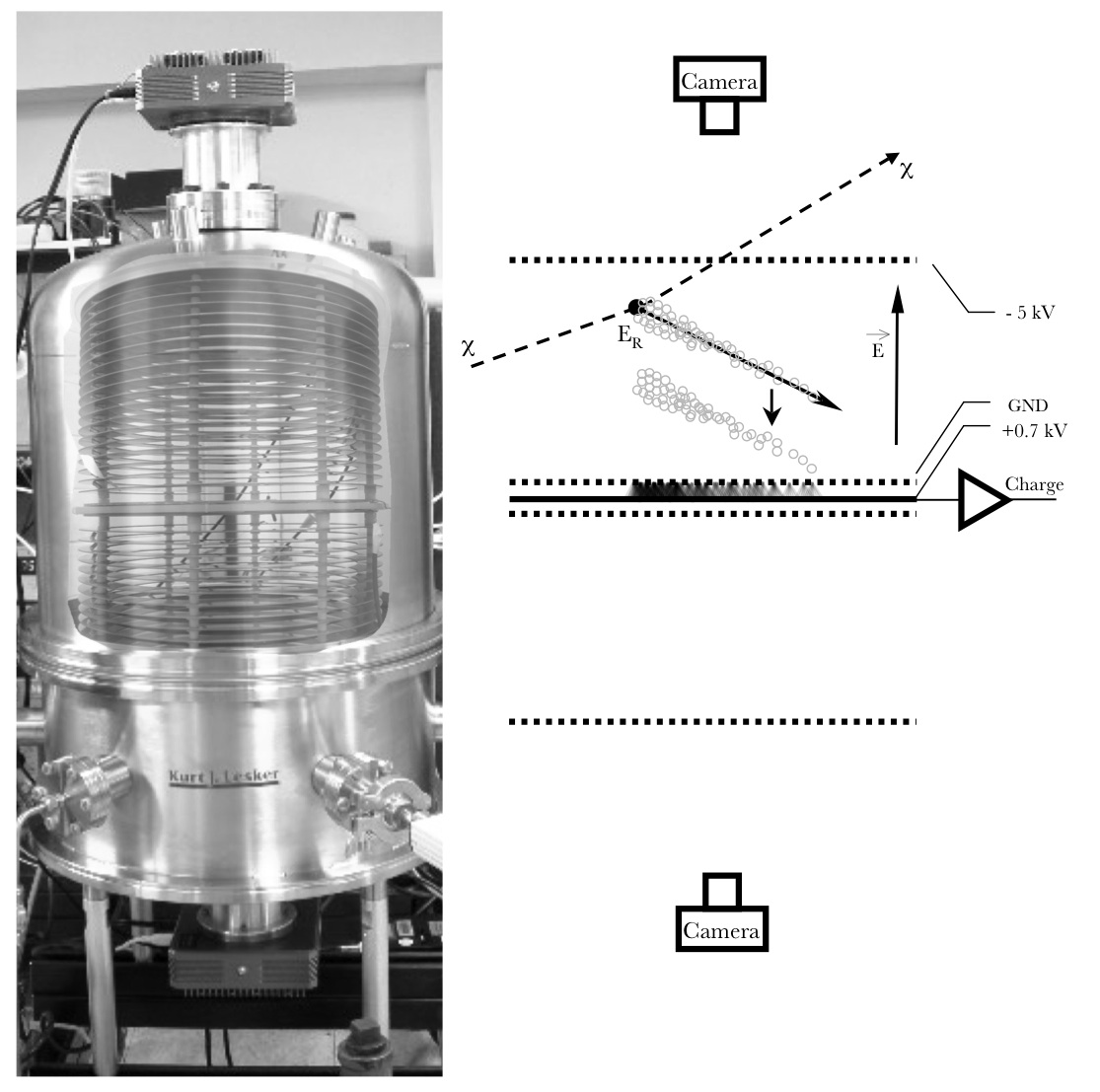}
\caption{(left) Photograph of the 10-liter DMTPC detector with an
  image of the dual TPC overlaid to provide an artificial glimpse
  inside the vacuum vessel.  The CCD cameras (top and bottom) each
  image an amplification region.  The stack of stainless steel field
  shaping rings condition the drift fields.  (right) A schematic
  representation of a WIMP-nucleus elastic scattering event in the
  detector.}
\label{fig:dmtpcDetector}
\end{figure}

The 10-liter DMTPC detector is shown in Fig.~\ref{fig:dmtpcDetector}.
The dual-TPC is housed inside a stainless steel vacuum vessel.  The
drift region is defined by a woven mesh cathode, typically at a
potential of -5~kV, separated from a wire mesh (28~$\mu$m wire,
256~$\mu$m pitch) ground grid 20~cm away.  The vertical drift field is
kept uniform to within 1\% by stainless steel field-shaping rings
spaced 1~cm apart.  An amplification region is formed between the
ground grid and a copper-clad G10 anode plane (at +720~V) which are
separated from each other by 500~$\mu$m using resistive spacers.  A
charge amplifier connected to the anode measures the ionization
generated by a particle moving through the detector.  A CCD camera
images the proportional scintillation light generated in the
amplification region.  The CCD camera and readout electronics are
located outside of the vacuum vessel.  The mesh-based amplification
region allows for two-dimensional images of charged particle tracks.

\begin{figure}\centering
\includegraphics[width=0.45\textwidth]{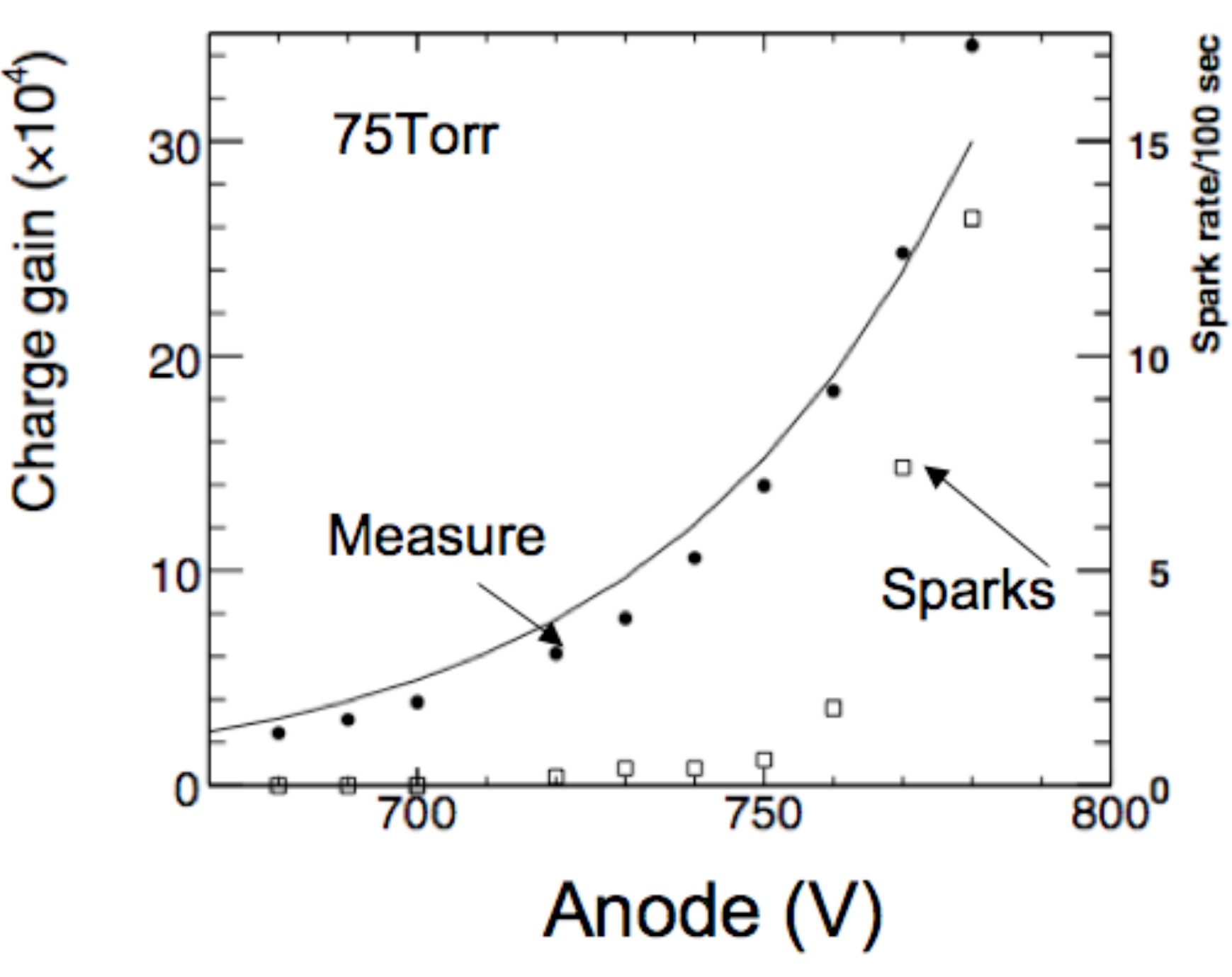}
\includegraphics[width=0.45\textwidth]{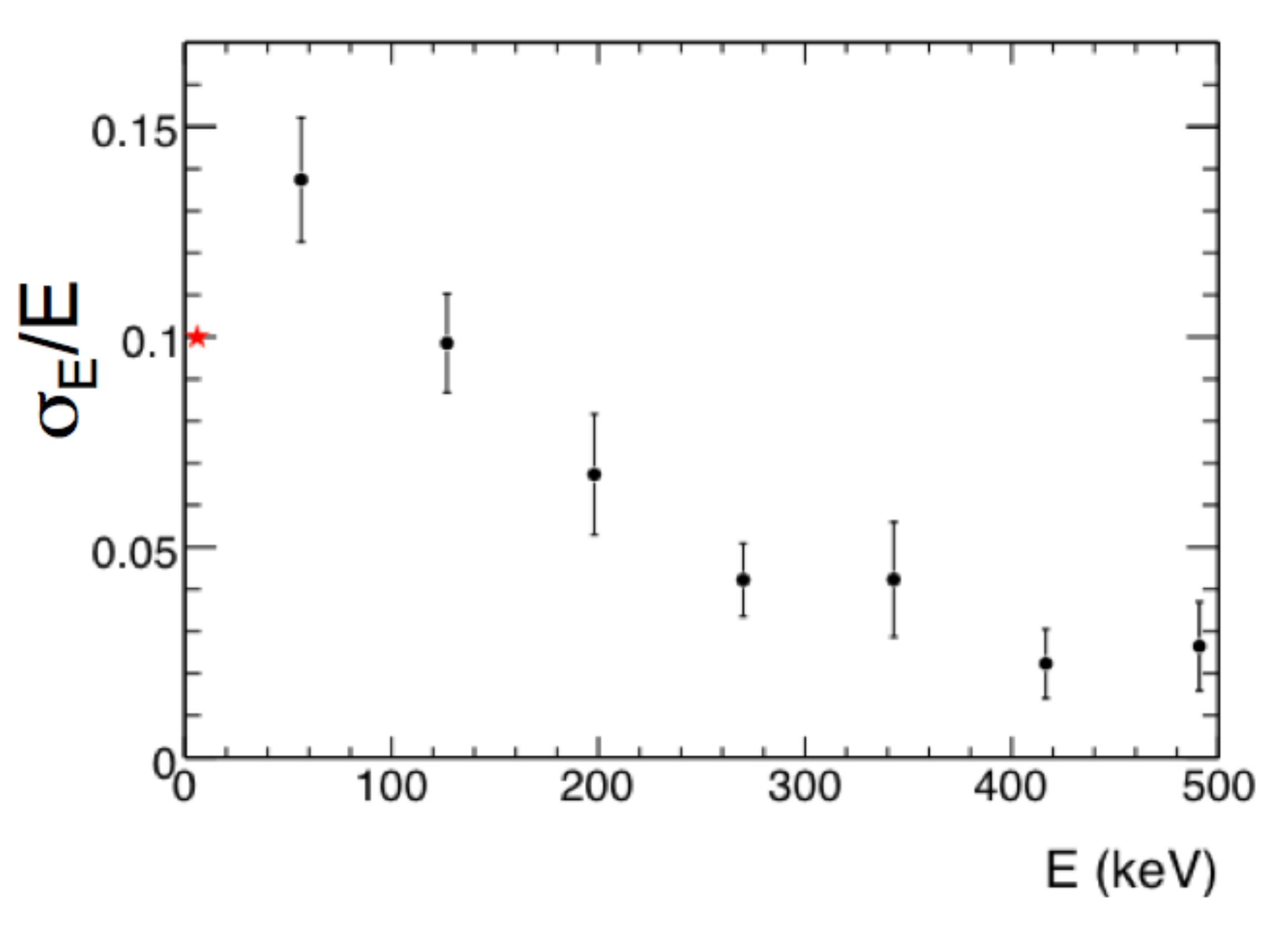}
\caption{(left) Gas gain and spark rate as a function of anode voltage for a mesh-based TPC filled with 75~Torr of \cffour.  The gas gain is measured from the charge collected from an $^{55}$Fe source.  (right) The energy resolution of the charge readout at 5.9~keV (star) and the CCD readout (circles) for energies above 50~keV.  }
\label{fig:dmtpcGainSpark}
\end{figure}

With a \cffour~pressure of 75~Torr, gas gains of $10^4-10^5$
are routinely achieved with minimal sparking (see
Fig.~\ref{fig:dmtpcGainSpark}).  The energy resolution of the charge
readout is $10$\% at 5.9~keV (measured with an $^{55}$Fe source), and
is $15$\% at 50~keV for the CCD readout (measured with an alpha
source, see Fig.~\ref{fig:dmtpcGainSpark}).  Since the stopping
$dE/dx$ in the detector is much smaller for electrons than for nuclear
recoils, the surface brightness of an electron track is dimmer, and
electron tracks are easily distinguished from nuclear recoils.  
The gamma rejection of our detector was measured to be $>10^6$ using an 8~$\mu$Ci $^{137}$Cs source \cite{dujmicTAUP2007}.

\subsection{\cffour~gas properties}
\cffour~has many advantages as a target gas for a dark matter search
in a gaseous TPC.  First, it has good sensitivity to spin-dependent
interactions because of its unpaired proton.  In addition, it is a
good counting gas, allowing gas gains in excess of $10^5$.  The
scintillation spectrum of \cffour~has significant emission in a broad
($\sim$100~nm wide) peak centered at 625~nm \cite{kabothCF4}. This
spectrum is well-matched to the peak quantum efficiency of CCDs.  In
addition, the measured transverse diffusion of electrons in \cffour~is
less than 1~mm for a 20~cm drift length at $E/N = 12\times 10^{-17}$~V
cm$^2$ (e.g. 300~V/cm at 75~Torr in the detector), and there is
negligible electron attachment over a 20~cm drift \cite{caldwell2009}.

\subsection{Head-tail measurements}
\begin{figure}\centering
\includegraphics[width=0.45\textwidth]{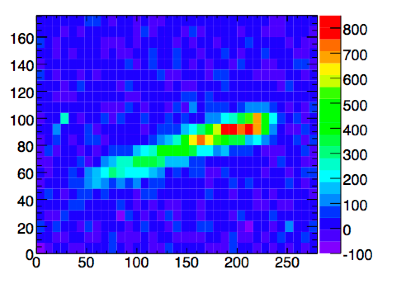}
\includegraphics[width=0.45\textwidth]{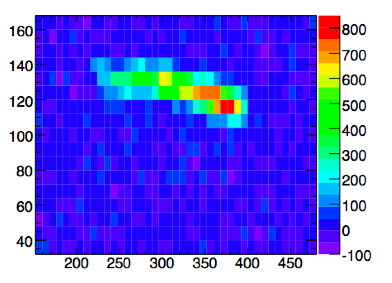}
\caption{Sample neutron-induced nuclear recoil candidates.  The neutrons were incident from the right.  The head-tail is evident from the light distribution along the track.  In these images, 100 pixels corresponds to 6~mm.}
\label{fig:dmtpcRecoils}
\end{figure}

As described in \cite{dujmicNIMA2008}, the DMTPC collaboration has
demonstrated the ability to measure the head-tail effect (the vector
direction of a recoil) on an event-by-event basis for energies down to
100~keV.  In that work, a $^{252}$Cf neutron source irradiated a
mesh-based detector filled with \cffour~at 75~Torr.  The CCD camera
acquired 6,000 one-second-exposure images, and 19 of these images
contained a candidate nuclear recoil.  Two examples of these
neutron-induced nuclear recoils are shown in
Fig.~\ref{fig:dmtpcRecoils}.  In these images, the nuclear recoil
axis and direction (head-tail) is clearly visible for each event.

\begin{figure}\centering
\includegraphics[width=0.45\textwidth]{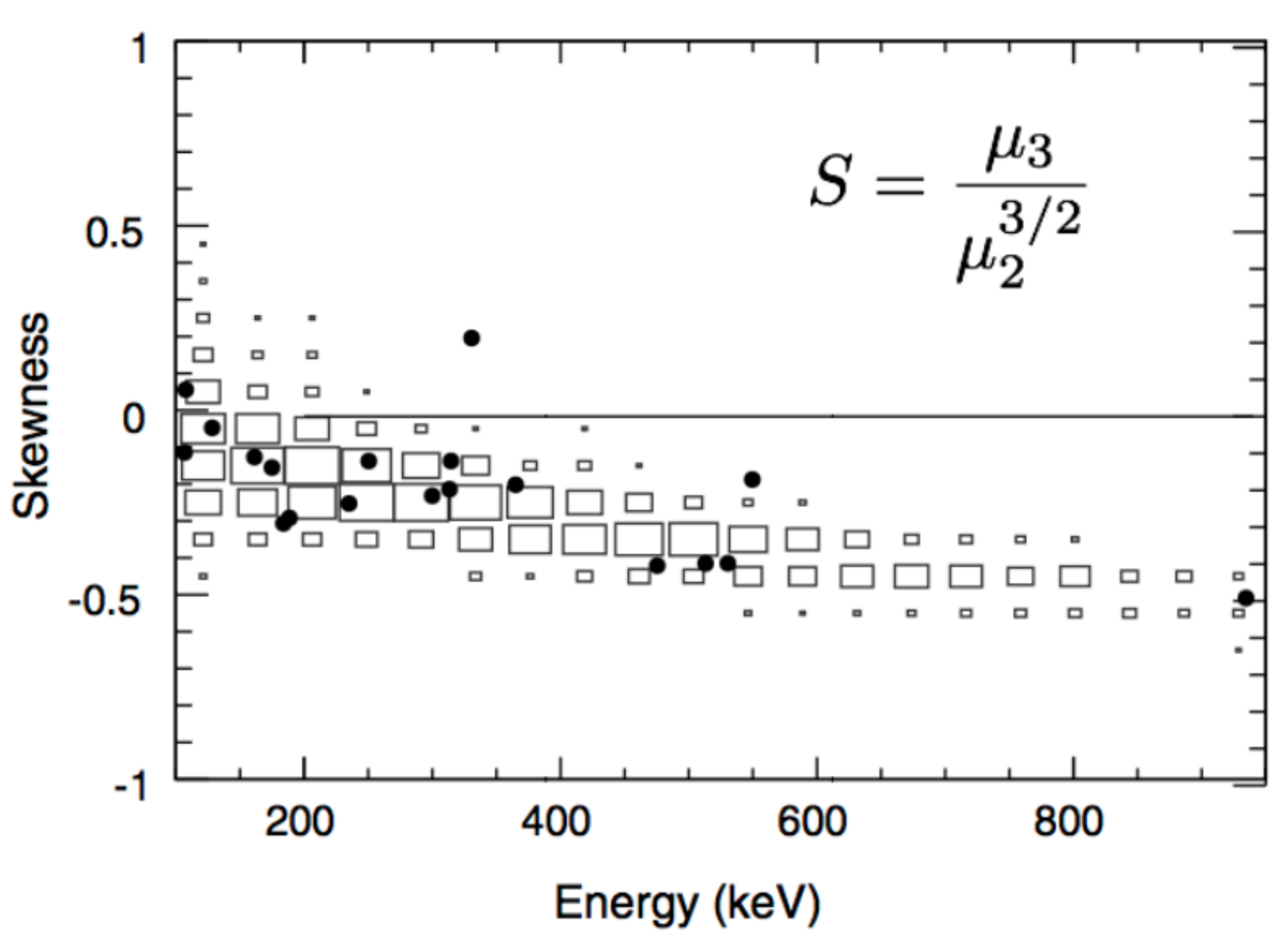}
\includegraphics[width=0.45\textwidth]{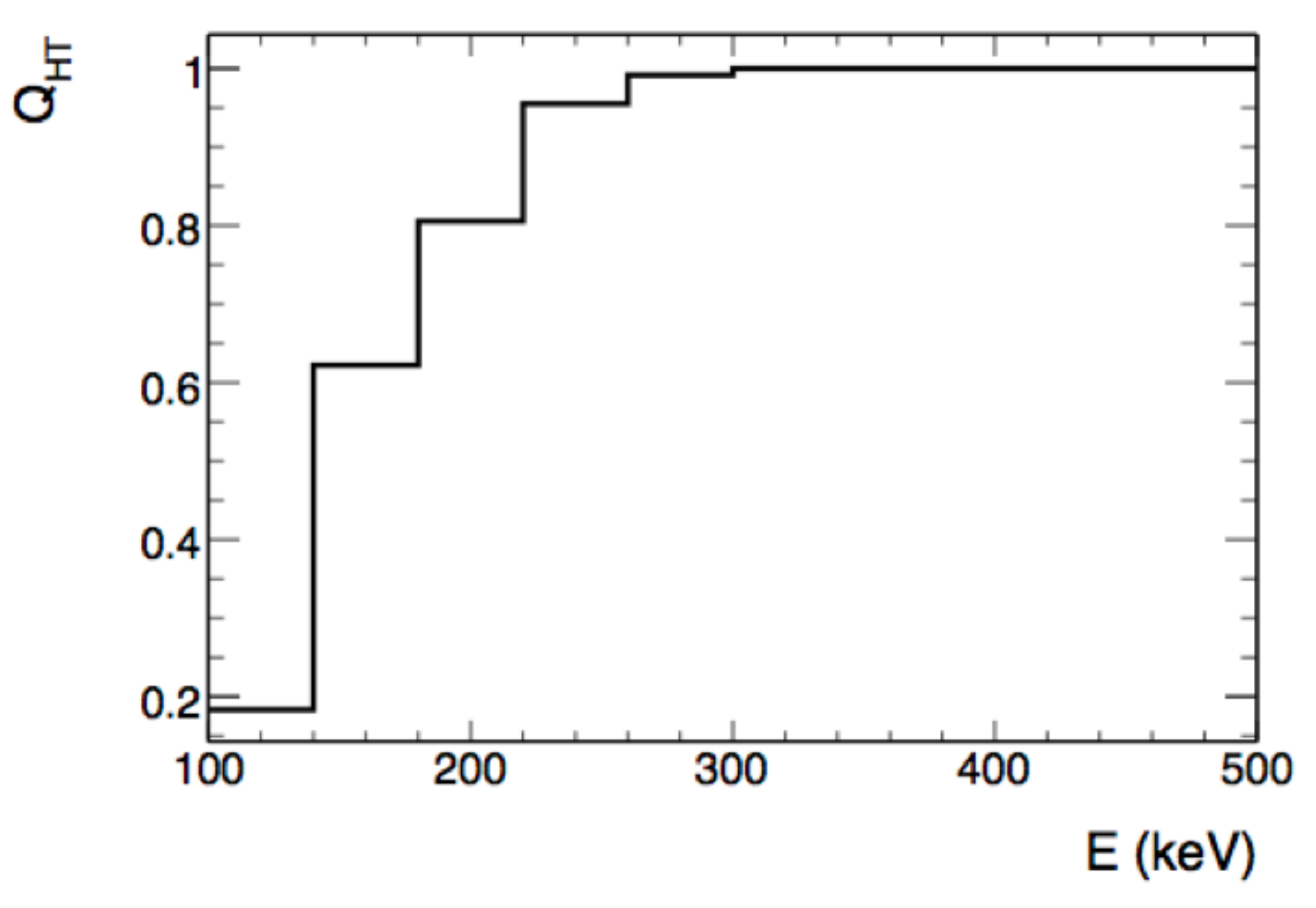}
\caption{((left) The observed (points) and predicted (box-histogram) skewness parameter, a measure of the head-tail effect, for candidate recoils induced by $^{252}$Cf neutrons in the detector.  A negative skewness parameter indicates that the head-tail was correctly reconstructed; (right) the head-tail quality factor $Q_{HT}$, as computed from Monte Carlo studies.}
\label{fig:dmtpcSkewnessQHT}
\end{figure}

Fig.~\ref{fig:dmtpcSkewnessQHT} shows the measured and
predicted range vs. energy for these events.  The recoil direction can
be measured from the light profile along the recoil track.  For the
candidate nuclear recoils, a dimensionless skewness parameter
$S=\mu_3/\mu_2^{3/2}$ is constructed, where $\mu_2$ and $\mu_3$ are
the second and third moments of the light distribution.  In our data
set, kinematics constrain all nuclei to be forward scattered and
therefore have negative skewness.
Fig.~\ref{fig:dmtpcSkewnessQHT} shows that the skewness can
be correctly reconstructed down to 100~keV.

For a set of nuclear recoils, the true forward-backward asymmetry is
$A=(F-B)/(F+B)$, where $F$ and $B$ are the number of forward and
backward recoils, respectively.  The measurement error on $A$ scales
like $\sigma_A\sim 1/\sqrt{N Q_{HT}}$, where $N$ is the total number
of measured recoils.  $Q_{HT}$ is a head-tail reconstruction quality
factor:
\begin{equation}
Q_{HT}(E_R)\equiv \epsilon(E_R) \left(\frac{N_{good}-N_{wrong}}{N_{good}+N_{wrong}}\right)^2
\end{equation}
where $E_R$ is the recoil energy, $\epsilon(E_R)$ is the (recoil energy
dependent) head-tail reconstruction efficiency, and $N_{good}$ and
$N_{wrong}$ are the number of events with head-tail correctly and
incorrectly reconstructed, respectively.  Monte Carlo studies show that $Q_{HT}$
exceeds 50\% above 140~keV (see Fig.~\ref{fig:dmtpcSkewnessQHT}).

\section{Surface Run Results}
\begin{figure}\centering
  \includegraphics[angle=90,width=0.45\textwidth]{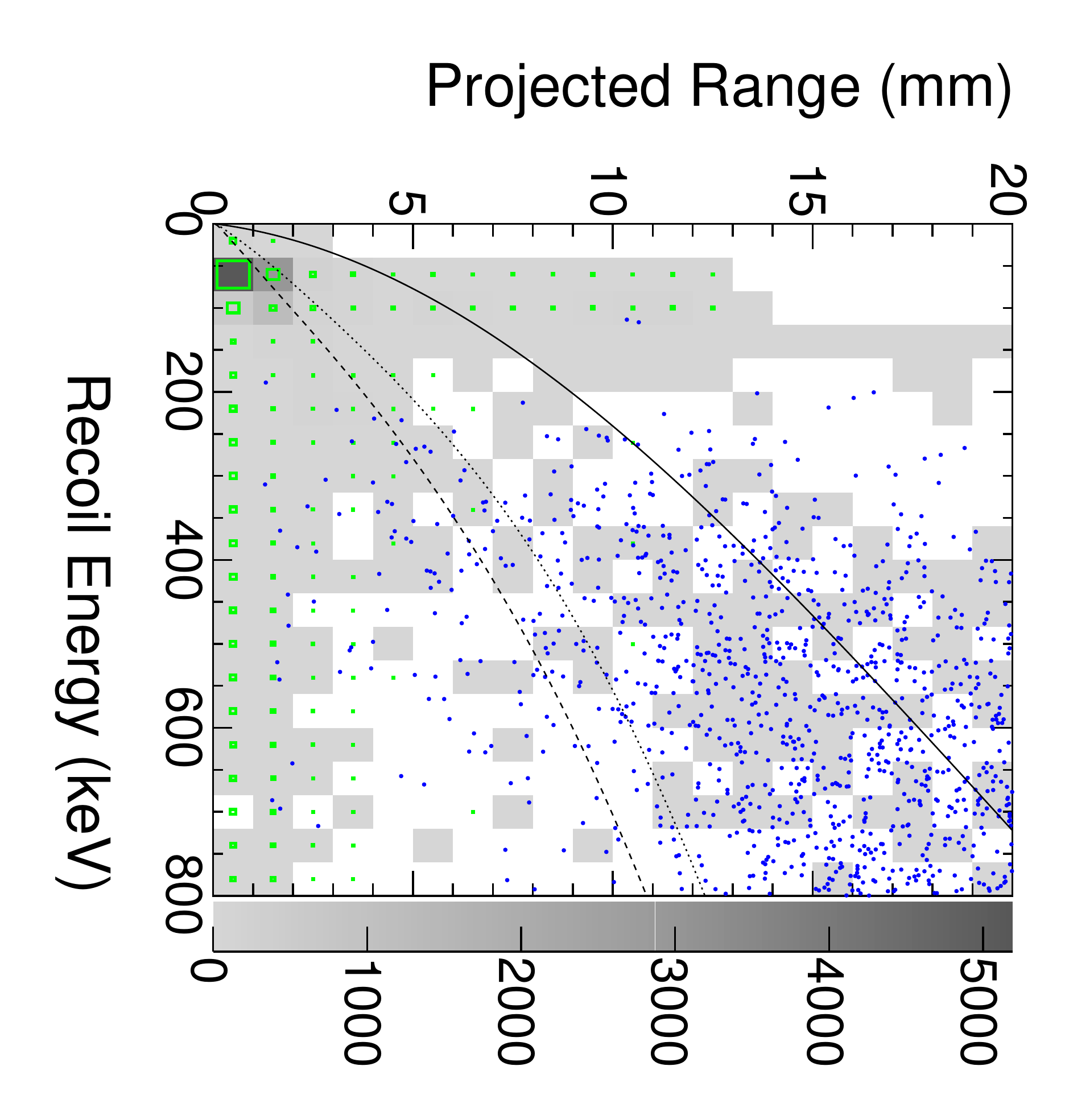}\\
  \includegraphics[angle=90,width=0.45\textwidth]{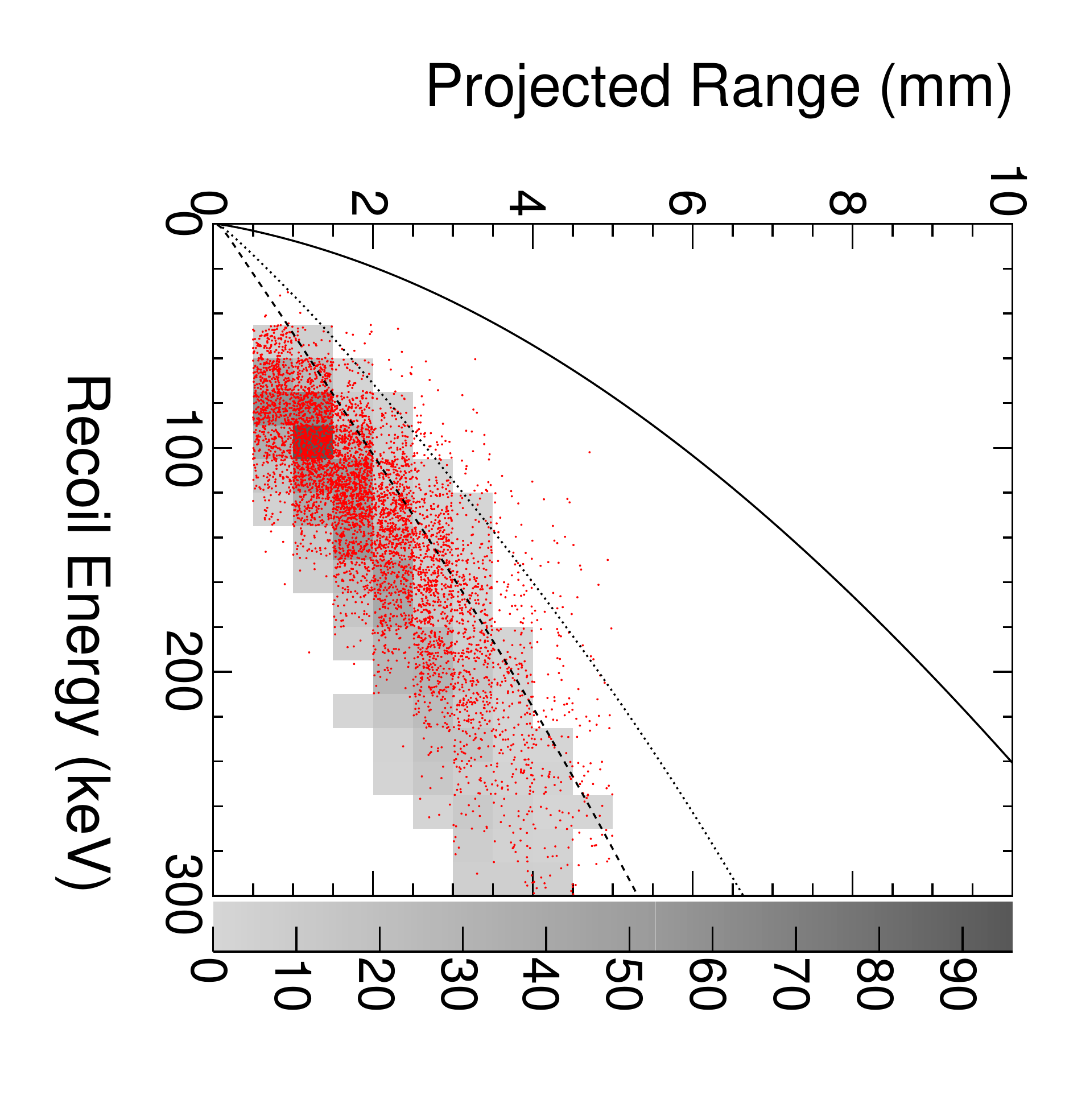}
  \includegraphics[angle=90,width=0.45\textwidth]{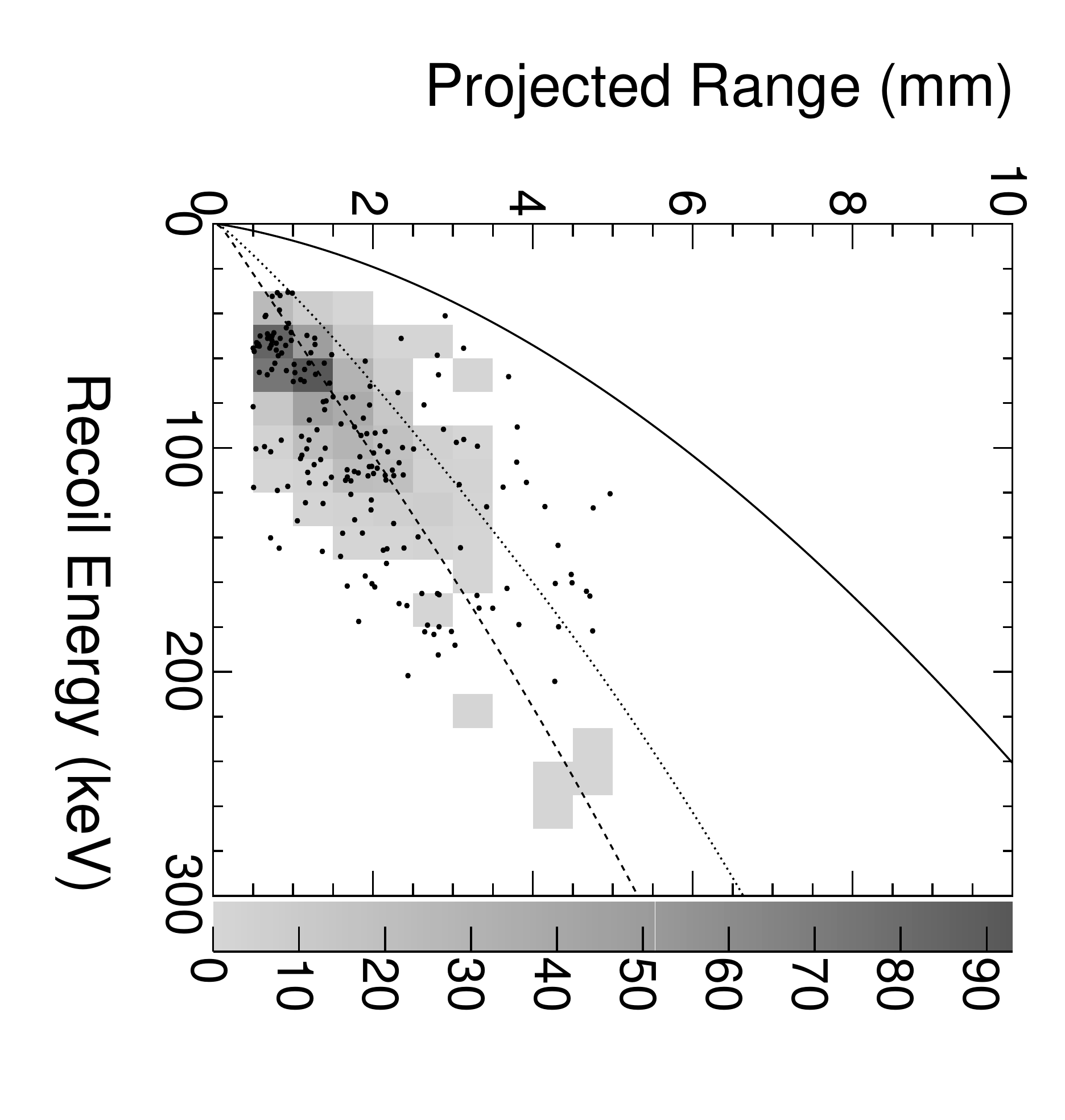}
  \caption{Reconstructed projected range vs. recoil energy. (top) Observed background populations in the surface run:  alphas (blue points), RBI (gray squares), and CCD interactions (green boxes).  See text for a description of these populations.  (bottom left) Calibration data (red points) from a $^{252}$Cf neutron source compared with Monte Carlo predictions (gray boxes).  The curves are the SRIM prediction for the 3D range of helium (solid), carbon (dotted) and fluorine (dashed) nuclei.  (bottom right) WIMP search data after nuclear recoil selection cuts (black points) compared with Monte Carlo predictions for a 200 GeV/c$^2$ mass WIMP.}
\label{fig:dmtpcWR2}
\end{figure}

The 10-liter detector was operated in a basement laboratory at MIT for an exposure of 35.7~g-day to study detector backgrounds prior to deploying the apparatus underground at WIPP.  Details of the experimental setup can be found in \cite{dmtpcSurfaceRun2010}.  Figure~\ref{fig:dmtpcWR2} (top) shows the events that were reconstructed in the camera images.  Over 99.9\% of these backgrounds are artifacts of the CCD camera and have no corresponding interaction in the active volume of the detector.  These backgrounds can be eliminated in the future by requiring coincident event detection in the CCD image and the charge readout (the charge signal was not available for the surface run).  The CCD backgrounds include direct interactions of charged-particles with the silicon CCD chip (referred to by astronomers as cosmic ray events) \cite{janesick}, and residual bulk images, a well-known effect in front-illuminated CCD cameras in which high-intensity long wavelength light \cite{janesick,rest} (e.g. from sparks in the amplification region) populates trapped states in the CCD chip that leak out with a thermal time-constant.  In the 10-liter data, RBI events can persist for 30~minutes or longer, but are easily identified (and cut) by their coincident position from frame to frame following a spark.

Using a $^{252}$Cf neutron source to irradiate the detector, and applying cuts on track topology, position, energy and range (for details about the image analysis, see \cite{dmtpcSurfaceRun2010}), we find that the remaining events follow the SRIM prediction for nuclear recoils.  The bottom left plot in Figure~\ref{fig:dmtpcWR2} shows the projected range of a track as a function of the recoil energy.  The lines are the SRIM predictions for the 3D range expected for (top to bottom) helium, carbon and fluorine nuclei.  Also plotted (red points) are the measured projected ranges from the neutron calibration data and (gray boxes) the Monte Carlo prediction, showing that the observed recoils populate the expected region of parameter space.  In the bottom right plot of Figure~\ref{fig:dmtpcWR2}, we show the events that pass our nuclear recoil selection cut, and compare this result with the Monte Carlo prediction.  The two agree well and are also consistent with the SRIM prediction for fluorine (and carbon) nuclear recoils.  

In an energy window of $80-200$~keV (chosen to maximize the integral above threshold of the efficiency and the expected recoil spectrum from a 200~GeV/c$^2$ WIMP), we find 105 remaining events.  Using the surface neutron flux measurement in \cite{neutrons}, we calculate an expected neutron background of 74 events, with a large uncertainty as the detector and building materials were not modeled in this calculation.  Taking all 105 events and computing the spin-dependent WIMP-proton cross-section 90\% confidence limit assuming zero expected backgrounds, we find $2.0\times10^{-33}$~cm$^2$ at a WIMP mass of 115~GeV/c$^2$.  The full cross-section limit is shown in Figure~\ref{fig:dmtpcSensitivity}.

\subsection{From surface run to underground}
\begin{figure}\centering
  \includegraphics[angle=90,width=0.5\textwidth]{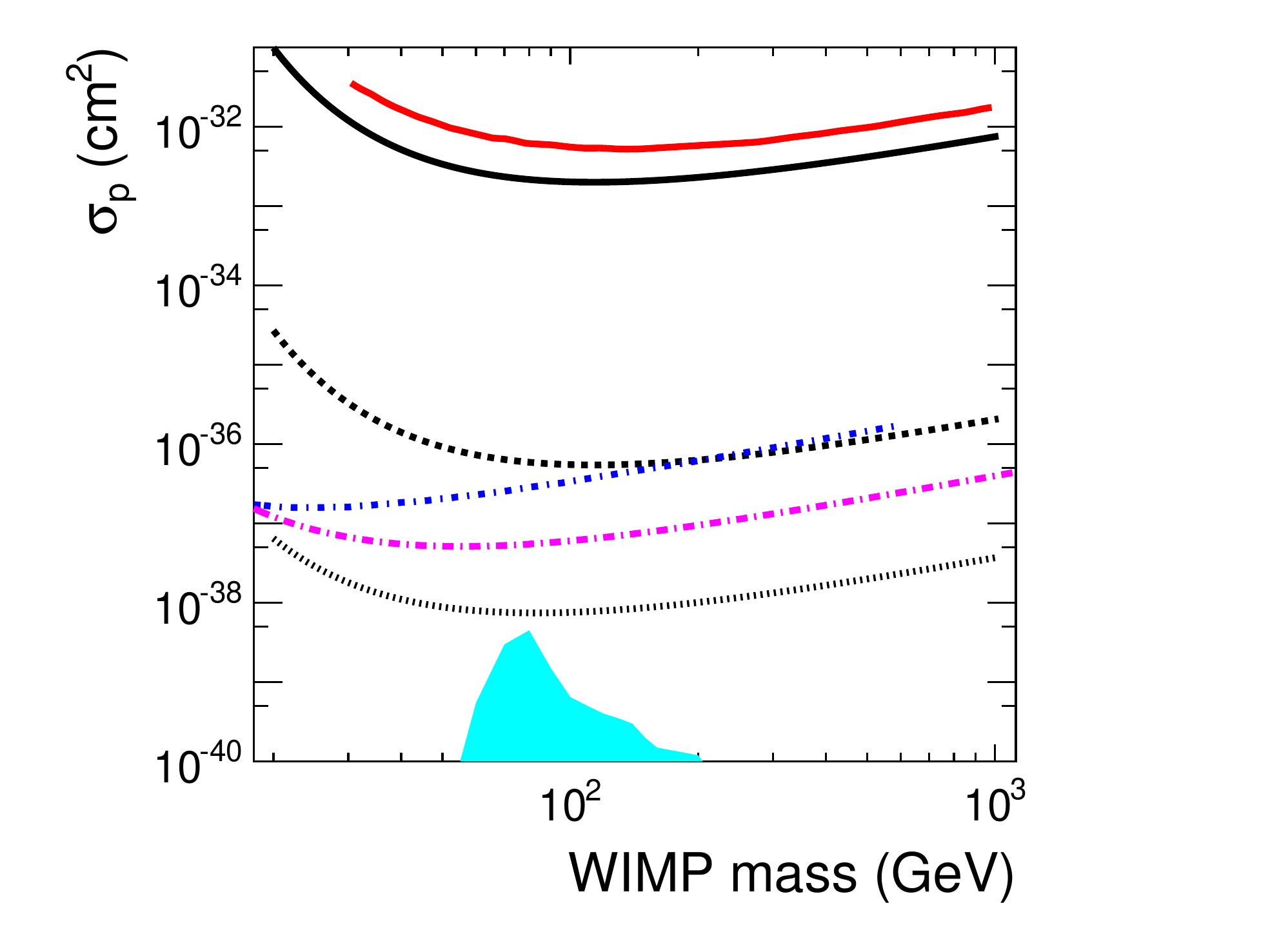}
\caption{Spin-dependent WIMP-proton cross-section limits and projected sensitivities.  The surface run limit from the DMTPC 10-liter run is shown as the solid black line.  Also shown is the surface run limit from the NEWAGE \cite{newage} directional experiment (solid red line), and limits from two non-directional experiments PICASSO (blue dash-dotted line) \cite{picasso} and COUPP (magenta dash-dotted line) \cite{coupp}.  Projected sensitivity curves are also shown for the DMTPC 10-liter detector at WIPP, run background-free for 1 year (black dashed line) and a 1~m$^3$ DMTPC detector at WIPP also run background-free for one year (black dotted line).  For the sensitivity of the 0.8~m$^3$ DRIFT directional detector, see their contribution in these proceedings.  The cyan shaded region shows MSSM parameter space \cite{mssm}.}
\label{fig:dmtpcSensitivity}
\end{figure}

The DMTPC collaboration has deployed the 10-liter detector underground
at WIPP (1.6~km.w.e. depth), where less than one neutron-induced event per year is expected.  In addition, we have built, and are commissioning at MIT, a 20-liter detector, made from highly radiopure materials and expect a significant reduction of alpha backgrounds.
DMTPC is also funded to construct a cubic meter detector.  When filled with 75~Torr of \cffour, the cubic meter detector will contain 380~grams of target material.  
The sensitivity of the cubic meter detector is shown in Figure~\ref{fig:dmtpcSensitivity}.

\end{document}